\documentstyle[preprint,tighten,epsfig,cite,aps]{revtex}
\newcommand{\elltbar}{\overline{\widetilde{\ell}\, }}
\newcommand{\eq}[1]{eq.~(\ref{#1})}
\newcommand{\rfn}[1]{(\ref{#1})}

\newcommand{\vev}[1]{\left\langle #1\right\rangle}

\begin{document}
\preprint{
\begin{tabular}{r}
FTUV--01--0802\\ 
IFIC/01--41 
\end{tabular}
}

\title{Neutrino masses from operator mixing}
\author{Josep F. Oliver and Arcadi Santamaria}
\address{\hfill \\ Departament de F\'{\i}sica Te\`orica and IFIC, 
Universitat de Val\`encia -- CSIC\\
Dr.\ Moliner 50, E-46100 Burjassot (Val\`encia), Spain}

\maketitle

\begin{abstract}
We show that in theories that reduce, at the Fermi scale, 
to an extension of the standard model with two doublets,
there can be additional dimension five operators giving rise to 
neutrino masses. In particular there 
exists a singlet operator which can not generate neutrino 
masses  at tree level but generates them through operator mixing. 
Under the assumption
that only this operator appears at tree level we calculate the neutrino
mass matrix. It has the Zee mass matrix structure and leads naturally to
bimaximal mixing. However, the maximal mixing prediction for solar neutrinos
is very sharp even when higher order corrections are considered. To allow for
deviations from maximal mixing a fine tuning is needed in the neutrino mass 
matrix parameters. However, this fine tuning relates the departure from 
maximal mixing in solar neutrino oscillations with the neutrinoless double 
beta decay rate.
\end{abstract}
\vspace{0.5cm}
\noindent PACS numbers: 14.60.Pq, 14.60.St, 14.80.Cp, 12.60.Fr\\
\vspace{0.5cm}

The simplest model for neutrino masses is based on the seesaw
mechanism\cite{Gell-Mann:1980vs,Yanagida:1980xy,Yanagida:1979}. In the 
seesaw mechanism the standard model (SM)
is enlarged with singlet right-handed neutrinos. Then, a Dirac mass 
term, $M_D$, mixing left-handed and right-handed neutrinos is possible.
In addition, since right-handed neutrinos do not carry any
gauge charge, they can have a Majorana mass, $M_R$, without compromising 
the gauge symmetry. If the right-handed Majorana mass term is very large,
as expected for a singlet mass term, very light Majorana neutrino masses for 
left-handed neutrinos are obtained through the diagonalization of the full 
mass matrix of neutral fermions, $m_\nu = M_D^2/M_R$, thus justifying the 
small size of neutrino masses. Since the Dirac mass term is proportional to 
the standard Higgs vacuum expectation value, this mechanism provides masses
which are $m_\nu \approx \Lambda_F^2/\Lambda$, being $\Lambda_F$ the 
Fermi scale and $\Lambda$ the lepton number breaking scale. This type
of behaviour is much more general and, in fact, 
many neutrino mass models 
can be cast into this form. 
This can be understood in the following way: if the 
SM is
just the low energy effective manifestation of some underlying theory, 
the effects of new physics can be represented by a series of gauge invariant
operators containing the SM fields with higher dimension operators suppressed 
by powers of the scale of new 
physics\cite{Weldon:1980gi,Weinberg:1979sa,Wilczek:1979hc}. At low energies, 
the most relevant 
operators will be those with the lowest dimension, that is dimension five 
operators. One can easily see that there is only one gauge invariant operator 
of dimension five one can build with the field content of the standard 
model\cite{Weinberg:1979sa}
 \begin{equation}
 \label{eq:seesaw1}
  {\cal L}_{\mathrm{seesaw}}  =  -\frac{1}{4} \frac{1}{\Lambda}
  (\elltbar F \vec{\tau} \ell)
  (\widetilde{\varphi}^\dagger \vec{\tau} \varphi) \ ,
 \end{equation}
where $\ell$ is the standard left-handed doublet of leptons,
$\widetilde{\ell} = i\tau_2 \ell^c$,  $\ell^c = C\overline{\ell}^T$
( $C$ is the charge conjugation operator),
$\varphi$ is
the Higgs doublet and $\widetilde{\varphi} = i\tau_2 \varphi^*$,
$\vec{\tau}$ are the Pauli matrices in $SU(2)$ space, $F$ is a
complex symmetric matrix in flavour space ($SU(2)$ and flavour indices have
been suppressed) and $\Lambda$ is a scale related to the scale of new physics.
It is clear that this Lagrangian does not conserve
generational lepton numbers,  but in addition it does not conserve the
total lepton number,  which is violated in two units. In the SM, lepton number
is conserved because the requirement of renormalizability and the small 
particle content (no right-handed neutrinos, no triplet scalars, etc) but as
long as the spectrum is enlarged there is no strong reason for lepton number 
conservation. Operator (\ref{eq:seesaw1}) makes this statement explicit. 
Therefore, this operator will be generated in any extension of the SM that 
does not conserve lepton number.

When the Higgs develops a vacuum expectation value (VEV),
operator (\ref{eq:seesaw1}) will give rise to a neutrino Majorana mass matrix 
given by
\begin{equation}
\label{eq:mnu}
 M_{\nu} = F \frac{v^2}{\Lambda} \ ,
\end{equation}
with $v=\vev{\varphi^{(0)}}=~174$~GeV, the SM Higgs VEV.
If we take the largest eigenvalue of  $F$ to be of order 1 and
use the laboratory bound on the
$\tau$-neutrino mass, $m_{\nu_{\tau}} < 18$~MeV, we find that  $\Lambda  >
10^6$~GeV.  Should one take the value suggested by atmospheric neutrino
data, $m_{\nu_{\tau}} \approx 0.06$~eV, one would obtain
$\Lambda \approx 5\times 10^{14}$~GeV, a scale which is close to the
unification scale: physics at very high energy scales could affect low energy
physics in a measurable way through neutrino masses. 
However the relationship between $\Lambda$ and masses of new particles 
can be quite different from the naive expectations. For instance, it seems
that the Lagrangian of \eq{eq:seesaw1} is generated by the exchange, 
among the leptons and the Higgses, of a scalar triplet, $\chi$, 
with hypercharge 1.  Then, $1/\Lambda\approx \mu/m^2_\chi$ with $\mu$ the 
trilinear coupling of the triplet with the Higgs doublets. However,
this is not the only possibility. In fact, \eq{eq:seesaw1} can be identically
rewritten, after a $SU(2)$ Fierz transformation, as
\begin{equation}
 \label{eq:seesaw2}
  {\cal L}_{\mathrm{seesaw}}  =  - \frac{1}{2}\frac{1}{\Lambda}
   (\elltbar \varphi) F  (\widetilde{\varphi}^\dagger \ell)
\ ,
\end{equation}
which suggests the exchange of a neutral heavy Majorana fermion; then
$\Lambda$ should be the mass of that fermion. Indeed, the seesaw 
mechanism described above implements naturally this possibility.

Given the full generality of this mechanism, which naturally relates the 
smallness of neutrino masses to the new physics scale, it is quite reasonable 
to try to fit the different neutrino mass models into this description. 
However, this is
not always possible: on one side it can happen that neutrino masses are
generated only by operators with higher dimensions (for a recent analysis of
the different possibilities see \cite{Babu:2001ex}) on the other side models
in which neutrino masses are generated through radiative corrections are 
also difficult to fit in the simplest scheme. 

In this paper, we want to show that the uniqueness of the effective
seesaw mechanism can be relaxed a bit if new fields are allowed at the 
electroweak scale, in particular in models with two light doublets.
This opens the door for a new class of effective seesaw mechanism in 
which the light neutrino masses are generated radiatively. This fact will 
allow us to lower the lepton number breaking scale by several orders of 
magnitude.  In addition,  as we will see, this mechanism predicts
a very particular form for the neutrino mass matrix which seems well
suited for explaining both atmospheric and solar neutrino data. 
This is not strange since a particular realization of this mechanism
is the Zee model\cite{Zee:1985rj,Zee:1980ai} and its variations. For instance,
models with spontaneous symmetry breaking of lepton number by a doublet 
\cite{Bertolini:1988kz}. Another interesting variation are models in which
spontaneous lepton number breaking is induced by a hyperchargeless 
triplet\cite{Santamaria:1989fh,Chang:1988aa}. They have the interesting 
property that although the triplet feels gauge interactions it does not 
contribute to  the invisible 
$Z$ decay width and that the lepton number breaking VEV could be at the 
electroweak scale (no bounds from red giant cooling by Majoron emission). 
Models with hyperchargeless triplets have also been discussed 
recently\cite{Forshaw:2001xq,Blank:1998qa} in connection with the results of 
last
standard model fits. Variations solving the strong CP problem can be found
in \cite{Arason:1991sg,Bertolini:1991vz}. All of these models predict a 
Zee type neutrino mass matrix (for a comprehensive review of extensions of
the Higgs sector of the SM see~\cite{Gunion:1989we,Gunion:1992hs}). 
Recent fits to neutrino 
data\cite{Bahcall:2001zu,Fogli:2001vr,Garzelli:2001zv,Barbieri:2000sv}
suggest some type of bimaximal mixing which can be 
accommodated naturally in this type of models 
\cite{Frampton:1999yn,Jarlskog:1998uf,Smirnov:1997bv,Smirnov:1994wj}. This
observation has boosted again the interest of models with Zee type neutrino 
matrix\cite{Koide:2001xy,Balaji:2001ex,Koide:2000jm,Chang:1999hg,Mclaughlin:1999rr,Joshipura:1999xe,Cheung:1999az}. 

In the same way that a variety of see-saw mechanism models can be described 
as a single effective operator it would be interesting to see 
if this class of models and perhaps other type of models with two light 
doublets can be described at low energies with just a few operators.
 
We will assume that the low energy (Fermi scale) theory is just the SM
model, with no right-handed neutrinos, supplemented by an additional doublet. 
We will denote the two doublets as $\varphi_1$ and $\varphi_2$. 
Then, in principle, nothing forbids that both 
doublets couple to the two types of quarks and to the leptons. However, 
it is well known that this, in general, will lead to neutral current 
flavour changing problems\cite{Glashow:1977nt}.
Therefore, for the moment we will consider that only one of the doublets, 
$\varphi_1$, couples to the fermions. This can be archived naturally 
by assigning an additional conserved charge to the doublet $\varphi_2$, 
lepton number for example. Of course, this additional charge should be 
explicitly broken in the Higgs potential in order to avoid the appearance of 
a Goldstone boson once the doublet $\varphi_2$ acquires a VEV (models in which
lepton number is broken spontaneously by a doublet 
\cite{Bertolini:1988kz,Santamaria:1987uq,Aulakh:1982yn} are excluded by the
invisible $Z$ width measured at LEP). Therefore, for the moment we will 
assume the SM Yukawa couplings to leptons
\begin{equation}
\label{eq:sm-yukawa}
{\cal L}_Y = \bar{\ell} Y e_R \varphi_1 + {\mathrm h.c.}
\end{equation}
Again, $SU(2)$ and flavour indices have been suppressed and $Y$ is 
a $3\times 3$ complex matrix in flavour space. However, because no right-handed
neutrinos have been introduced in the model and because only 
$\varphi_1$ couples
to leptons, it can be chosen, without loss of generality, as diagonal with
all its elements real and positive. As in the SM, neutrinos remain massless
because there are no right-handed neutrinos and because lepton number is 
conserved. 

Now we will assume that this model is just the low energy manifestation of
a more complete theory which will only shows up at higher energies. 
As discussed 
above, if the scale of new physics is high enough its effects
can be be parametrized by operators with higher dimensionality: 
\begin{equation}
\label{eq:opexpansion}
{\cal L}_{eff}= {\cal L}_{\mathrm 2HSM} + \frac{1}{\Lambda} {\cal L}_1 +
                        \frac{1}{\Lambda^2}{\cal L}_2+ \cdots\ .
\end{equation}
Here ${\cal L}_{\mathrm 2HSM}$ represents the renormalizable Lagrangian
we just outlined which is a minimal extension of the SM containing two 
doublets. 
 
If the theory at the Fermi scale contains two doublets, one can write four 
independent dimension five operators\footnote{In general one also expects 
higher dimension operators which could contribute to interesting 
processes as $\mu$-$e$
conversion in nuclei, $\mu\rightarrow e \gamma$, etc. For instance, 
when a charged scalar is integrated out all those operators appear at
one loop\cite{Bilenkii:1994bt}.}
\begin{equation}
 \label{eq:triplet-bis}
 {\cal L}_{T_1}  =  -\frac{1}{8} \frac{1}{\Lambda}
  (\elltbar T_1 \vec{\tau} \ell)
  (\widetilde{\varphi_1}^\dagger \vec{\tau} \varphi_1),\quad 
 {\cal L}_{T_2}  =  -\frac{1}{8} \frac{1}{\Lambda}
  (\elltbar T_2 \vec{\tau} \ell)
  (\widetilde{\varphi_2}^\dagger \vec{\tau} \varphi_2)~,
\end{equation} 
\begin{equation}
\label{eq:triplet}
  {\cal L}_{T}  =  -\frac{1}{4} \frac{1}{\Lambda}
  (\elltbar T \vec{\tau} \ell)
  (\widetilde{\varphi_2}^\dagger \vec{\tau} \varphi_1)~,
\end{equation}
and
\begin{equation}
\label{eq:singlet}
  {\cal L}_{S}  =  -\frac{1}{4} \frac{1}{\Lambda}
  (\elltbar S \ell)
  (\widetilde{\varphi_2}^\dagger \varphi_1) \ .
 \end{equation}

Operators ${\cal L}_{T_1}$, ${\cal L}_{T_2}$ can be excluded by the same
symmetry used to forbid Yukawa couplings of the doublet $\varphi_2$ to the 
fermions. For instance, one can assign lepton number to $\varphi_2$ in such
a way that it does not have Yukawa coupling to fermions while allowing
the couplings ${\cal L}_{T}$ and ${\cal L}_{S}$. $L(\varphi_2)=-2$ will do 
the job and will forbid operators  ${\cal L}_{T_1}$, ${\cal L}_{T_2}$
Therefore, in the following
we will only consider operators  ${\cal L}_{T}$ and ${\cal L}_{S}$. 

It is important to notice that the operator in (\ref{eq:singlet}) does not 
exist with only one doublet because the singlet coupling of two scalar 
doublets is antisymmetric (in $SU(2)$ components it is just 
$\epsilon_{ij} \varphi_{2i} \varphi_{1j}$). In addition,  
one can see that, due to Fermi statistics, the $3\times 3$ matrix in 
flavour space, $T$, is 
symmetric while the singlet coupling $S$ is a complex antisymmetric matrix. 
 
\begin{figure}[htbp]
\begin{center}
\vspace{1cm}
\epsfig{file=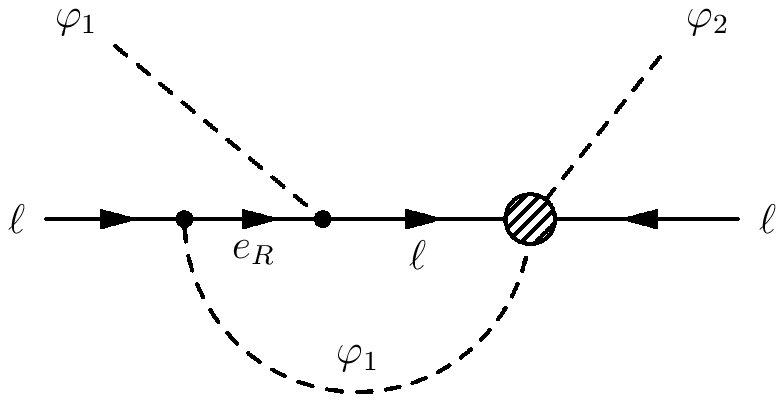}
\end{center}
\vspace{1cm}
\caption{\label{fig:fig1}Diagram contributing to the mixing of the singlet and
triplet operators}
\end{figure}

When the Higgs fields develop a charge conserving VEV, operator
${\cal L}_{T}$ gives rise to a neutrino
Majorana mass. However, the singlet operator ${\cal L}_{S}$ 
does not give any mass to the neutrinos because the product of the two 
doublets has hypercharge 1, and a singlet with hypercharge 1  has also 
charge 1 (at difference with the triplet combination (\ref{eq:triplet}) 
which has neutral components). For this 
reason the singlet operator does not seem very interesting at first sight. 
However, a very interesting situation arises when, for some reason 
(limited particle content of the full theory), only operator 
(\ref{eq:singlet}) arises at tree level\footnote{An example of this situation 
is provided by the Zee model in which a charged singlet, $h^+$, and an extra 
doublet are introduced with couplings $\elltbar f \ell h^+$ and 
$\mu h^- \tilde{\varphi}^\dagger_2 \varphi_1$, then, for a heavy $h^+$ one 
finds, at tree level, that $S/\Lambda = 4\mu f /m^2_h$ while the triplet 
operator cannot be 
obtained at tree level.}. As commented before, it cannot give rise to neutrino
masses after spontaneous symmetry breaking.  However, one expects that 
renormalization effects will mix all operators with the same dimensionality and
the same quantum numbers. Therefore, (\ref{eq:singlet}) and (\ref{eq:triplet}) 
will mix under the renormalization group and operator (\ref{eq:triplet}) will
be generated at one loop even if it did not appear at tree level.  
In fact, one can easily see, by computing the diagram
in fig.~\ref{fig:fig1} (and the crossed diagram) and taking the divergent 
part, that the matrix $T$ obeys the following renormalization group equation 
(RGE)
\begin{equation}
\label{eq:rge}
\mu \frac{d}{d\mu} T = 
\frac{1}{(4\pi)^2} \left(S Y Y^\dagger+Y^* Y^T S^T\right) + \cdots
\end{equation}   
Here, the dots represent extra contributions to the renormalization 
group of the $T$ matrix which are proportional  to the $T$ matrix itself and,
therefore, cannot generate it if it did not exist at some scale. 
This RGE is peculiar with respect the RGE we are used to see for the
Yukawa couplings, in both the SM and the MSSM, in what there is a piece that
does not depend on the $T$ matrix. In both the SM and the MSSM there are
several chiral symmetries broken only by the Yukawa couplings that ensure
that the RGE of those couplings should transform in a covariant way with
respect to those symmetries. This ensures, in this type of theories,
that fermion masses cannot be generated through radiative corrections.
Equation~(\ref{eq:rge}) is not of this type 
and, therefore, even if the coupling $T$ did not exist at the scale $\Lambda$
at which the operators were generated, it will arise through operator 
mixing. It is very easy to integrate \eq{eq:rge} by keeping
only the leading logarithm (a more sophisticated integration can be
performed by taking into account the running of the standard Yukawa
couplings, the running of the $S$ coupling itself and the extra couplings
we neglected in \eq{eq:rge}, however, this effect is higher order in the 
couplings and since the couplings are small we expect a small effect). 
The result is that 
\begin{equation}
\label{eq:tmz}
T(m_Z) \approx 
\frac{1}{(4\pi)^2} \left(S Y Y^\dagger+Y^* Y^T S^T\right)\log
\left(\frac{m_Z}{\Lambda}\right)+T(\Lambda)~, 
\end{equation} 
where we have identified $m_Z$ with the Fermi scale. We will assume that
$T(\Lambda)$ is not generated at tree level.
Of course $T(\Lambda)$ could
also pick up contributions at one loop (or higher loops). To compute
them one would need to know the details of the full theory in which our 
effective theory is embedded. However if $\Lambda$ is
large enough $T(m_Z)$ will be dominated by the model independent logarithmic
piece in \eq{eq:tmz} which can be computed in the effective theory. 
So, as a first approximation we will assume $T(\Lambda)\approx 0$ and further
on we will keep $T(\Lambda)$ only when the logarithmic pieces vanish.

After spontaneous symmetry breaking, if both $\varphi_1$ and $\varphi_2$
develop a VEV, operator~(\ref{eq:triplet}) will give rise to a neutrino mass
matrix for the lefthanded neutrinos given by
\begin{equation}
\label{eq:numass}
M_\nu \approx \tan\beta \frac{1}{(4\pi)^2} 
\left(S Y Y^\dagger+Y^* Y^T S^T\right)
\log\left(\frac{m_Z}{\Lambda}\right) \frac{v_1^2}{\Lambda}~,
\end{equation}
where $v_1=\vev{\varphi^{(0)}_1}$, $v_2=\vev{\varphi^{(0)}_2}$ are the
VEV's of the two doublets and $\tan\beta = v_2/v_1$. 
The seesaw structure is apparent in the last term. The other
factors, however, are also important. First, the neutrino mass matrix
comes naturally proportional to the mass of the leptons squared which gives 
an important suppression since lepton Yukawa couplings are small.
Second, it contains the standard loop suppression factor
$1/(4\pi)^2$ and, third, the ratio of VEV's of the second doublet $v_2$ to the 
standard doublet VEV, $\tan \beta = v_2/v_1$, can give 
an additional suppression factor. All together, one can achieve with this
mechanism the same neutrino masses one would obtain in a standard
seesaw mechanism with a $\Lambda$ which is at least $6$ orders of magnitude
smaller. This could put the scale of the new physics responsible 
for neutrino masses at the reach of
the next generation of accelerators if the ratio of VEV's $v_2/v_1$  
and/or the largest $S_{ij}$ are very small. 
However, perhaps the most interesting aspect of \eq{eq:numass} is the 
structure of the mass matrix, inherited
from the antisymmetric structure of the singlet coupling $S$: if we choose
for the Yukawa coupling of the leptons a diagonal form we find
\begin{equation}
\label{eq:numass2}
M_\nu \approx -m_0 \left(
\begin{array}{ccc}
0 & S_{e \mu} (x_e-x_{\mu}) & S_{e\tau} (x_e-1) \\
S_{e \mu} (x_e-x_{\mu}) & 0 & S_{\mu\tau} (x_{\mu}-1) \\ 
S_{e \tau} (x_e-1) & S_{\mu\tau} (x_{\mu}-1)  & 0
\end{array}
\right)~,
\end{equation}
with $x_e= m^2_e/m^2_{\tau}$, $x_{\mu} = m^2_{\mu}/m^2_{\tau}$ and
\[
m_0 = \tan\beta 
\frac{m^2_{\tau}}{\Lambda (4\pi)^2}\log\left(\frac{\Lambda}{m_Z}\right)~.
\]
The structure of this mass matrix is very interesting since all diagonal
elements are zero and contains only three free parameters (the three 
elements of the antisymmetric matrix $S$).
It is convenient to rewrite it as
\begin{equation}
\label{eq:numass3}
M_\nu \approx \left(
\begin{array}{ccc}
0 & m_{e\mu} & m_{e\tau} \\
m_{e\mu} & 0 & m_{\mu\tau} \\ 
m_{e\tau} & m_{\mu\tau} & 0
\end{array}
\right)~.
\end{equation}
This form of mass matrix has been considered recently in order to fit 
both atmospheric and solar neutrino data\cite{Frampton:1999yn,Jarlskog:1998uf,Smirnov:1997bv,Smirnov:1994wj}.
Let us review some of the results: in order to explain solar neutrino
data in terms of oscillations one needs a mass squared difference, $\Delta_s$ 
which is 
$10^{-11} {\mathrm eV}^2 \leq \Delta_s \leq 10^{-5} {\mathrm eV}^2$. A global 
analysis including the results of SNO suggests mixings close to maximal.
On the other hand, in order to explain atmospheric neutrino data one needs 
a mass  squared difference,
$\Delta_a\approx 3\times 10^{-3} {\mathrm eV}^2$ and also a very large mixing. 
The particular structure of the obtained mass matrix (it is traceless) 
implies that the sum of the eigenvalues
is zero a fact which constraints the possible solutions. An analysis of the
different possibilities in terms of this mass matrix has been carried out in
\cite{Frampton:1999yn,Jarlskog:1998uf,Smirnov:1997bv,Smirnov:1994wj} where it 
has been shown that only the case with $m_{e\mu} \sim m_{e\tau}$
and $m_{\mu\tau}\ll m_{e\mu}, m_{e\tau}$ is acceptable. This naturally 
predicts maximal mixing for solar oscillations which, after SNO, seems to be 
the only viable possibility. 
In fact the Zee mass matrix predicts, in this case\cite{Koide:2001xy},
\begin{equation} 
\label{eq:sin2t}
\sin^2 2\theta_s = 1-\frac{1}{16}\left(\frac{\Delta_s}{\Delta_a}\right)^2 ~.
\end{equation}
This is a very strong prediction which is compatible with the
LOW and the VAC solutions of the solar neutrino problems. The LMA solution,
which right now seems to be the preferred solution, would be marginally 
compatible with this prediction. However, it is important to notice that 
in general we expect 
corrections to the Zee mass formula. By requiring that only one doublet 
couples to fermions we have taken the most restrictive (and predictive) 
possibility. By relaxing this assumption one can perfectly fit also
the LMA region in solar neutrino parameters \cite{Balaji:2001ex}. But even in 
the restrictive case in which only one doublet couples to fermions one
can also take into account subdominant contributions. In fact, on general
grounds we expect modifications to the particular form of the mass matrix
with all the elements in the diagonal vanishing. There is no symmetry that
enforces this structure, therefore one expects that at some point this
structure will receive corrections. This happens, for instance in the Zee
model where diagonal entries are generated at two loops \cite{Chang:1999hg}. 
For our purposes we can include all those corrections in the initial 
contributions at the scale $\Lambda$, that is, by considering a non-vanishing
$T(\Lambda)$ which we could take as a general symmetric matrix. Given the 
bimaximal mixing required by the data it is natural to parametrize the neutrino
mass matrix as follows
\begin{equation}
\label{eq:mnu-perturbation}
M_\nu \approx 
m_0 \left[\left(
\begin{array}{ccc}
0 & \frac{1}{\sqrt{2}} & \frac{1}{\sqrt{2}} \\
\frac{1}{\sqrt{2}} & 0 & 0 \\ 
\frac{1}{\sqrt{2}} & 0 & 0
\end{array}
\right)+
\epsilon \left(
\begin{array}{ccc}
\gamma_{ee} & \gamma_{e\mu}  & \gamma_{e\tau} \\
\gamma_{e\mu} & \gamma_{\mu\mu} & \gamma_{\mu\tau} \\ 
\gamma_{e\tau} & \gamma_{\mu\tau} & \gamma_{\tau\tau}
\end{array}
\right)\right]~,
\end{equation}
where, without loss of generality we can take $\gamma_{e\tau}=-\gamma_{e\mu}$ 
and choose one of the $\gamma$'s equal to one defining in this way the
normalization of $\epsilon$. We will assume that the first term, generated
through running and containing the logarithmic enhancement, is the dominant 
one and gives the scale of atmospheric neutrinos.
The term proportional to $\epsilon$ is subdominant and cannot be computed in
the effective theory. Its explicit form depends on the details of the 
underlying theory and it could contain one loop contributions not enhanced
by large logarithms or/and higher loop contributions, depending on the 
underlying theory. For simplicity, we will also take 
all the $\gamma$'s as real.
Notice that the mechanism suggested in
this paper only gives a leading order contribution with zeros in the diagonal.
The extra structure assumed in the leading mass term (
$m_{e\mu} \approx m_{e\tau}$ and $m_{\mu\tau} \approx 0$) must be imposed
with some additional symmetries which, however, are not unnatural in this
type of models \cite{Zee:1985rj}. 
The modifications introduced by the term proportional to $\epsilon$ can
be qualitatively very important. One of the most interesting predictions
of the leading order mass matrix is that there is no neutrinoless double 
beta decay (NDBD) : since the NDBD  amplitude is proportional
to $(M_\nu)_{ee}$ it is obvious that a mass matrix with zeroes in the diagonal
forbids NDBD. However any correction to the leading
order mass matrix introducing diagonal entries will lead to NDBD. On the
other hand,
corrections to the leading order mass matrix are necessary in order to 
accommodate solar neutrino mass differences and small departures from maximal 
mixing in solar and/or atmospheric neutrinos.   
It is important to check how these corrections could modify the sharp 
predictions of the model.      

By diagonalizing the mass matrix at second order in $\epsilon$ one easily
obtains that
\begin{eqnarray}
\label{eq:dmsolar}
\Delta_a & \approx & m^2_0~, \label{eq:delta-atm}\\
\sin^2 2\theta_a & \approx & 1-
\frac{1}{2}
\left(16\gamma_{e\mu}+(\gamma_{\mu\mu}-\gamma_{\tau\tau})^2\right)\epsilon^2 ~,\label{eq:sin-atm}\\
\Delta_s &\approx& m^2_0(2\gamma_{\mu\tau}+2\gamma_{ee}+\gamma_{\mu\mu}+
\gamma_{\tau\tau})\epsilon ~,\label{eq:delta-solar}\\
\sin^2 2\theta_s &\approx & 1
-\frac{1}{16}\left(\frac{\Delta_s}{\Delta_a}\right)^2 +
\frac{1}{2}\left(
\gamma_{ee}(2\gamma_{\mu\tau}+\gamma_{\mu\mu}+\gamma_{\tau\tau})-
(\gamma_{\mu\mu}-\gamma_{\tau\tau})^2\right)\epsilon^2 ~.\label{eq:sin-solar}
\end{eqnarray}
Since $\Delta_s/\Delta_a$ is known to be small, the natural prediction of the
model is just maximal mixing for the two angles to a very good degree of
precision. Departures from maximal mixing are allowed naturally for the
atmospheric mixing angle since $\Delta_s$ does not depend on 
$\gamma_{e\mu}$ which controls the atmospheric mixing (for 
$\gamma_{\mu\mu}=\gamma_{\tau\tau}$), therefore it can be made large 
without any conflict with solar neutrino data. However, to accommodate 
deviations from maximal mixing
in the solar neutrino parameters is more delicate. One needs to make 
$\Delta_s$ small while keeping the last term in 
eq.~\rfn{eq:sin-solar} relatively large. 
This is clearly unnatural. For instance, in the case in which 
$\gamma_{\mu\mu} = \gamma_{\tau\tau}=0$ one obtains 
$\Delta_s/\Delta_a \approx 2\left(\gamma_{\mu\tau}+\gamma_{ee}\right)$ while
$\sin^2 2\theta_s \approx 1-(\Delta_s/\Delta_a)^2/16+
\gamma_{\mu\tau}\gamma_{ee}$. Therefore, to obtain a very 
small $\Delta_s/\Delta_a$
while keeping a sizeable contribution to $\sin^2 2\theta_s$ one would need to
fine tune the couplings in such a way that 
$\gamma_{ee} \approx -\gamma_{\mu\tau}$. Although it is not natural,
 this possibility might be interesting because perhaps it is the only way 
to accommodate LMA within this scheme and because if it is realized 
in nature it will link
the deviation from maximal mixing in solar neutrino oscillations with 
neutrinoless double beta decay: as commented before, the NDBD amplitude 
is proportional to the so-called effective neutrino mass 
$\langle m_\nu \rangle\equiv \sum U^2_{ei} m_{\nu i} = (M_\nu)_{ee}$.

In this work we have shown that in models that contain in its low-energy 
(Fermi scale) effective theory two Higgs doublets there are four independent
dimension five gauge invariant operators which violate lepton number. Three 
of them, which couple leptons to doublets in the triplet channel, generate
masses at tree level when the doublets acquire VEV's. The other operator, 
which couples leptons and doublets in the singlet channel, cannot 
generate masses at three level. However, loop corrections mix all operators
under the renormalization group and, therefore, the singlet operator also
gives rise to neutrino masses when the doublets acquire VEV's.

Under the assumption that only the singlet operator is generated at tree level,
at the scale of new physics, one can compute the induced neutrino masses at
one loop. Neutrino masses are suppressed by several factors, the loop factor,
the masses of charged leptons and the ratio of the two doublets VEV's, 
therefore allowing for a new physics scale 
several orders of magnitude lower than
the one needed in tree-level mechanisms for neutrino masses.

Because the structure of the singlet effective operator, which necessitates
antisymmetric couplings in flavour, the obtained mass has the structure
of the Zee mass matrix with zero entries in the diagonal. This structure
naturally accommodates bimaximal mixing and is well suited for explaining both
solar and atmospheric neutrino data. However, in the simplest scheme it is
very difficult to accommodate the LMA solution for solar 
neutrinos because it gives a sharp prediction for $\sin^2 2\theta_s = 1$ once
one takes into account the mass differences needed for solar neutrinos. 

We have investigated the possibility to accommodate 
$\sin^2 2\theta_s \not= 1$
in this scheme by considering non-leading contributions to the mass matrix 
since, on general grounds, one expects to generate non-zero entries for the
diagonal elements of the mass matrix. We found that the prediction 
$\sin^2 2\theta_s = 1$ is quite stable and, only by fine tuning the parameters
of the mass matrix, it is possible to accommodate at the same time 
$\sin^2 2\theta_s \not= 1$ and $\Delta_s/\Delta_a \ll 1$. This fine tuning,
even though unnatural, has the interesting property that relates the rate
of neutrinoless double beta decay to the departure from maximal mixing in
solar neutrino oscillations.
 
\vspace{0.5cm}

This work has been funded by CICYT under the Grant AEN-99-0692, 
by DGESIC under the Grant PB97-1261 and by the DGEUI of 
the Generalitat Valenciana under the Grant GV98-01-80.


\end{document}